\documentclass[a4paper, 10pt, oneside]{article}

\usepackage{jheppub}
\usepackage{feynmp}
\usepackage{graphicx}
\usepackage{graphics}
\usepackage{multirow}

\newcommand\fverb{\setbox\fverbbox=\hbox\bgroup\verb}
\newcommand\fverbdo{\egroup\medskip\noindent%
            \fbox{\unhbox\fverbbox}\ }
\newcommand\fverbit{\egroup\item[\fbox{\unhbox\fverbbox}]}
\newbox\fverbbox

\title{Lepton flavor violating decays of mesons to lepton-pairs in a gauge group $SU_{L}(2)\times
U_{Y_{1}}(1)\times SU_{X}(2)$}
\author{Fayyazuddin}
\affiliation{National Center for Physics and Physics Department,
Quaid-I-Azam University,Islamabad }
\emailAdd{fayyazuddins@gmail.com}

\date{\today}

\abstract{ The lepton flavor conserving and lepton flavor violating
decays of $K$ and $B$ mesons to lepton pairs in the gauge group
$G=SU_{L}(2)\times U_{Y_{1}}(1)\times SU_{X}(2)$ are discussed. The
quark-lepton transitions mediated by the lepto-quark bosons
$X_{\mu}^{\pm2/3}$ of the group $SU_{X}(2)$ provide a framework to
construct an effective Hamiltonian for these decays. The effective
coupling constant
$\frac{G_{X}}{\sqrt{2}}=\frac{g^{2}_{X}}{8m^{2}_{X}}$; $m_{X}$ is
the mass at which the group $G$ is broken to the SM group. The upper
bound on ($G_{X}/G_{F}$)is obtained from the most stringent
experimental limit on the $B.R(K^{0}_{L}\to \mu^{\mp}e^{\pm})$.
Several cases of pairing three generations of leptons and quarks in
the representation $(2,\bar{2})$ of the group are analyzed. For some
pairing, the upper bound on $(G_{X}/G_{F})$ is of the order
$(6-9)\times 10^{-6}$ and is compatible with the upper limits on
various LF violating $K$-decays . In particular for these cases, we
find the upper limit on the branching ratio
$B.R(K^{0}_{L}\to\mu^{-}\mu^{+})\sim (1.9-8.3)\times10^{-9}$. It is
shown that for LF violating $B$-decays to lepton pairs, the time
integrated decay rate
$B^{0}_{d,s}\to\ell^{-}\tau^{+}(\tau^{-}\ell^{+})$ is a promising
area to test the model.}

\begin{document}
\maketitle

\section{Introduction}\label{Intro}

In weak decays, mediated by charged vector bosons $W_{\mu}^{\pm}$,
the hadronic flavor changes but lepton flavor is conserved. The
relative strength of weak decays in hadronic sector is determined by
\text{CKM} matrix. However, the weak neutral current is flavor
conserving. The flavor changing neutral current \text{(FCNC)}
induced weak decays are not allowed at tree level. These decays are
highly suppressed and occur through loop diagrams in the standard
model. \text{FCNC} induced weak decays are lepton flavor conserving
in the standard model; there are stringent experimental limits on
the lepton flavor changing \text{(LFC)} decays.

The electromagnetic unification scale is \text{260} \text{Gev}. The
Higgs mass cannot be predicted in the electroweak theory
spontaneously broken by the scalar fields. The discovery of a new
particle \text{125} \text{Gev}, most likely to be Higgs boson has
implication for the "hierarchy problem". The "Higgs mechanism" of
the electroweak set the scale for all known particles. The
"hierarchy problem" is why it is so small compared to another mass,
the Planck mass ($10^{19}$ \text{}Gev) which is the fundamental unit
of mass in theory of gravitation.

It is of interest to explore the intermediate mass scales, between
the Higgs mass and the Planck mass. One way to do it is to extend
the electroweak group to a higher group \cite{[1]}; the extra vector
boson of the extended group provide the intermediate mass scales.

In this paper, we consider extension of the group $SU_L(2)\times
U_Y(1)$ to $SU_L(2)\times U_{Y_1}(1)\times SU_X(2)$. The three
generations of left-handed fermions
\begin{eqnarray}
\left(\begin{array}{cc} u_i & \nu_i\\
d^{\prime}_i & e_i\end{array}\right)_L,\quad i&=&e,\mu,\tau
(\nu_e,\nu_{\mu},\nu_{\tau})\notag\\
&=& d, s, b (u, c, t)\notag
\end{eqnarray}
($i$: is the generation index; the color index is suppressed) are
assigned to the representation
\begin{eqnarray}
(2, \bar 2)_{Y_1}:\quad {Y_1}&=&0\quad \text{for leptons}\notag\\
&=-&{2/3}\quad \text{for quarks}\notag
\end{eqnarray}
Note that $I_{X_3}=1/2,-1/2$ for the quarks and lepton respectively.
The charge $Q$ is given by
\begin{eqnarray}
Q={1\over 2}(\tau_{L_3}+\tau_{X_3}+Y_1)={1\over
2}(\tau_{L_3}+Y)\notag
\end{eqnarray}
where
\begin{align}
Y&=\tau_{X_3}+Y_1\notag\\
&=\begin{array}{rcl} -1 & \mbox{for leptons}\\
1/3 & \mbox{for
quarks}\notag
\end{array}
\end{align}
as in the standard model. The right-handed fermions are singlet:
\begin{eqnarray}
(1,1): Y&=&Y_1\notag\\
&=&{4/3}\quad \text{for up quarks}\notag\\
&-&{2/3}\quad \text{for down quarks}\notag\\
&-&2\quad \text{for charged leptons}\notag\\
&&0\quad \text{for neutrinos}\notag
\end{eqnarray}
The left-right symmetric gauge group $SU_L(2)\times SU_R(2)\times
U_Y(1)$ likewise can be extended.
\begin{eqnarray}
\left(\begin{array}{cc}u_i & \nu_i\\
d^{\prime}_i & e_i\end{array}\right)_L:(2,1,\bar
2)_{Y_1}\notag\\
\left(\begin{array}{cc}u_i & \nu_i\\
d^{\prime}_i & e_i\end{array}\right)_R:(1,2,\bar{2})_{Y_1}\notag
\end{eqnarray}
Note that
\begin{eqnarray}
\left(\begin{array}{c}d^{\prime}\\s^{\prime}\\b^{\prime}\end{array}\right)&=&
V\left(\begin{array}{c}d\\s\\b\end{array}\right)\notag
\end{eqnarray}
where $V$ is the usual \text{CKM} matrix.

A salient feature of the model is quark-lepton transitions mediated
by the lepto-quarks $X_{\mu}^{-2/3},X_{\mu}^{2/3}$ carrying baryon
and lepton numbers $(1/3,-1),(-1/3,1)$ respectively. These
transitions generate the flavor changing current, containing both
lepton flavor conserving and lepton flavor violating. We first
consider the group $SU_L(2)\times U_{Y_1}(1)\times SU_X(2)$; the
left-right symmetric model is discussed in the last section.

\section{Interaction Lagrangian}

It is straight forward to write the Lagrangian. The interaction
Lagrangian is given by
\begin{eqnarray}
L_{int}=&-&g\sin\theta_W J^{\mu}_{em}A_{\mu}-{g\over
2\sqrt2}\big[(\bar\nu_i\Gamma^{\mu}_Le_i+\bar u_i\Gamma^{\mu}_L
d_i^{\prime})W_{\mu}^{+}+h.c\big]\notag\\
&-&{g\over\cos\theta_W}J^{Z\mu}Z_{\mu}-\frac{g_X}{\sqrt{1-g^{\prime
2}/g_X^2}}J^{Z^{\prime}\mu}Z^{\prime}_{\mu}\notag\\
&-&{g\over
2\sqrt2}\Big\{\sum_i\sum_j\big[\bar\nu_i(DU)^{\dag}_{ij}\Gamma^{\mu}_L
u_j+\bar e_i(DU)^{\dag}_{ij}\Gamma^{\mu}_L
d^{\prime}_j\big]X_\mu^{-2/3}+h.c\Big\}\label{1}
\end{eqnarray}
where $\Gamma_L^\mu=\gamma^\mu(1-\gamma^5),J^{Z\mu}$ is the weak
neutral current coupled to $Z_{\mu}$ and
\begin{eqnarray}
J^{Z^{\prime}\mu}={1\over4}&&\Big[(\bar u_i\Gamma^{\mu}_L u_i+\bar
d_i\Gamma^{\mu}_Ld_i -\bar
e_i\Gamma^{\mu}_Le_i-\bar\nu_i\Gamma^{\mu}_L\nu_i)\notag\\
&&-{g^{\prime 2}\over g_X^{2}}\big(4J^{\mu}_{em}-(\bar
u_i\Gamma^{\mu}_L u_i-\bar d_i\Gamma^{\mu}_Ld_i
+\bar\nu_i\Gamma^{\mu}_L\nu_i-\bar
e_i\Gamma^{\mu}_Le_i)\big)\Big]\label{2}
\end{eqnarray}
If we take $g_X=g$,
\begin{eqnarray}
\frac{g_X}{\sqrt{1-g^{\prime
2}/g_X^2}}=\frac{g}{\sqrt{1-\tan^{2}\theta_W}}\label{3}
\end{eqnarray}
\begin{align}
J^{Z^{\prime}\mu}=\tan^{2}\theta_W J^{Z\mu}-{1\over
4}\Big[4\sin^2\theta_W J^{\mu}_{em}&-(\bar u_i\Gamma^{\mu}_L
u_i+\bar
d_i\Gamma^{\mu}_Ld_i)\notag\\&+(\bar\nu_i\Gamma^{\mu}_L\nu_i+\bar
e_i\Gamma^{\mu}_Le_i)\Big]\label{4}
\end{align}
The matrix $D$ in the last term of Eq.(\ref{1}) is the $3\times 3$
matrix associated with the permutation group $S_3$ :
\begin{eqnarray}
D(e)=D(123)=\left(\begin{array}{ccc}1 & 0 & 0\\0 & 1 & 0\\0 & 0 &
1\end{array}\right)\notag
\end{eqnarray}
\begin{eqnarray}
D(13)=\left(\begin{array}{ccc}0&0&1\\0&1&0\\1&0&0\end{array}\right),\quad
D(23)=\left(\begin{array}{ccc}1&0&0\\0&0&1\\0&1&0\end{array}\right)\notag\\
D(231)=\left(\begin{array}{ccc}0&1&0\\0&0&1\\1&0&0\end{array}\right),\quad
D(12)=\left(\begin{array}{ccc}0&1&0\\1&0&0\\0&0&1\end{array}\right)\label{5}
\end{eqnarray}
$D(13)$ interchanges first and third generation, $D(23)$
interchanges second and third generation, $D(12)$ interchanges first
and second generation and $D(231)$ interchange $(123)\rightarrow
(231)$ of leptons respectively in the representation $(2,\bar 2)$. U
in Eq.(\ref{1}) is a $3\times 3$ unitary matrix for lepton mixing.
In introducing these matrices, we have utilized the freedom of
permutation and mixing of three generations of leptons in the
representation $(2,\bar{2})$. For $D=D(e)$ and $U=1$, we have the
standard assignment \emph{viz}:
\begin{equation}
\left(\begin{array}{cc} u & \nu_e\\
d\prime & e
\end{array}\right)_L,
\quad \left(\begin{array}{cc} c & \nu_\mu\\
s\prime & \mu \end{array}\right)_L, \quad \left(\begin{array}{cc} t
& \nu_\tau\\
b\prime & \tau
\end{array}\right)_L\notag
\end{equation}

The physical neutral vector bosons $A_{\mu}$, $Z_{\mu}$ and
$Z^{\prime}_{\mu}$ are related to neutral vector bosons $W_{3\mu}$,
$B_{1\mu}$ and $X_{3\mu}$ of the extended group as follows
\begin{eqnarray}
W_{3\mu}=\frac{1}{\sqrt{g^2+g^{\prime
2}}}(g^{\prime}A_{\mu}+gZ_{\mu})\notag\\
X_{3\mu}=\frac{1}{\sqrt{g_1^{2}+g_X^{2}}}(g_1B_{\mu}+g_XZ^{\prime}_{\mu})\notag
\end{eqnarray}
\begin{subequations}
\begin{eqnarray}
B_{1\mu}=\frac{1}{\sqrt{g_1^{2}+g_X^{2}}}(g_XB_{\mu}-g_1Z^{\prime}_{\mu})\label{6-a}
\end{eqnarray}
\text{where vector boson}
\begin{eqnarray}
B_{\mu}=\frac{1}{\sqrt{g^2+g^{\prime
2}}}(gA_{\mu}-g^{\prime}Z_{\mu})\label{6-b}
\end{eqnarray}
\end{subequations}
is coupled to the Abelian group $U_Y(1)$ of the standard model. The
gauge coupling $g^{\prime}$ of $U_Y(1)$ and the electromagnetic
coupling are related to the gauge couplings $g$, $g_1$ and $g_X$ as
follows
\begin{eqnarray}
{1\over g^{\prime 2}}&=&{1\over g^{2}_1}+{1\over g^{2}_X}\notag\\
{1\over e^{2}}&=&{1\over g^{2}}+{1\over g^{\prime 2}}\label{7}
\end{eqnarray}

The gauge group $SU_L(2)\times U_{Y_1}(1)\times SU_X(2)$ is
spontaneously broken to the group $SU_L(2)\times U_Y(1)$ by
introducing a scalar doublet $\eta$ of $SU_X(2)$:
\begin{eqnarray}
\eta=(\eta^{2/3}, \eta^{0}); Y_1=1/3\Big(1-{2\over 3}\Big), 1\notag
\end{eqnarray}
The relevant term in the Lagrangian of the scalar doublet $\eta$:
\begin{eqnarray}
\Big[&\partial^{\mu}\eta+{i\over
2}g_X\eta\overrightarrow{\tau_X}.\overrightarrow{X}^{\mu}+{i\over
2}g_1\eta
B^{\mu}_1-{i\over2}{2\over3}g_1\eta^{2/3}B^{\mu}_1\Big]\notag\\
\times\Big[&\partial_{\mu}\bar\eta-{i\over
2}g_X\overrightarrow{\tau_X}.\overrightarrow{X}_{\mu}\bar\eta-{i\over
2}g_1
B_{1\mu}\bar\eta+{i\over2}{2\over3}g_1B_{1\mu}\eta^{-2/3}\Big]\notag
\end{eqnarray}
Now $\eta$ can be put in the form:
\begin{align}
\eta=(\eta^{2/3},
\eta^{0})=&\left(\eta^{2/3},\frac{\text{v}^{\prime}+H_X+ih_X}{\sqrt2}\right)\notag\\
\rightarrow&\left(0,\frac{\text{v}^{\prime}+H_X}{\sqrt2}\right)\notag
\end{align}
by spontaneous symmetry breaking the group $SU_L(2)\times
U_{Y_1}(1)\times SU_X(2)$ to the standard model group $SU_L(2)\times
U_Y(1)$. Note that the scalars $\eta^{2/3}$ and $h_X$ have been
absorbed to give masses to $X_{\mu}^{\pm 2/3}$ and $Z^{\prime}$
respectively. The vector bosons mass term is given by
\begin{align}
&-{1\over4}\left(0,\frac{\text{v}^{\prime}+H_X}{\sqrt2}\right)\Big(g_X\overrightarrow{\tau_X}.\overrightarrow{X}^{\mu}
+g_1B_1^{\mu}\Big)\Big(g_X\overrightarrow{\tau_X}.\overrightarrow{X}_{\mu}
+g_1B_{1\mu}\Big)\left(\begin{array}{cc}0\\
\frac{\text{v}^{\prime}+H_X}{\sqrt2}\end{array}\right)\notag\\
&=-{1\over4}\left(\frac{\text{v}^{\prime}+H_X}{\sqrt2}\right)^2\Big[2g_X^{2}X^{2/3
\mu}X_{\mu}^{-2/3}+(g_1^{2}+g_X^{2})Z^{\prime\mu}Z^{\prime}_{\mu}\Big]\notag
\end{align}
Hence we have
\begin{eqnarray}
m_X^{2}={1\over4}g_X^{2}\text{v}^{\prime2}\label{8}
\end{eqnarray}
\begin{eqnarray}
m^{2}_{Z^{\prime}}&=&{1\over4}(g_1^{2}+g_X^{2})\text{v}^{\prime2}={1\over4}\frac{g_X^{2}\text{v}^{\prime2}}{(1-g^{\prime
2}/g_X^2)}\notag\\
&=&\frac{m_X^{2}}{(1-g^{\prime 2}/g_X^2)}\label{9}
\end{eqnarray}
For $g_X=g$,
\begin{eqnarray}
m_X^{2}&=&{1\over4}g^{2}\text{v}^{\prime2}=m_W^{2}\frac{\text{v}^{\prime2}}{\text{v}^{2}}\notag\\
m^{2}_{Z^{\prime}}&=&\frac{m_X^{2}}{(1-\tan^2\theta_W)}\label{10}
\end{eqnarray}
By the symmetry breaking mechanism discussed above, the standard
model is decoupled from the extended group at a mass scale $m_X$.
The vector bosons $X_{\mu}^{\pm 2/3}$ and $Z^{\prime}_\mu$ acquire
super heavy masses and give the intermediate mass scale. The
stringent experimental limits on lepton flavor violating decays
generated by the exchange of bosons $X_{\mu}^{\pm 2/3}$ provide a
lower bound on the intermediate mass scale.

\section{Effective Hamiltonian for FCNC}

The flavor changing current coupled to $X_{\mu}^{\pm 2/3}$ are given
by
\begin{eqnarray}
J^{X\mu}&=&\bar
e_{i}(U^\dag)_{ij}(D^\dag)_{jk}\Gamma^{\mu}_{L}d^{\prime}_{k}\notag\\
&=&\bar
e_{i}(U^\dag D^\dag)_{ik}V_{k\alpha}\Gamma^{\mu}_{L}d_{\alpha}\notag\\
&=&\bar e_{i}(U^\dag D^\dag
V)_{i\alpha}\Gamma^{\mu}_{L}d_{\alpha}=C_{i\alpha}\bar
e_{i}\Gamma^{\mu}_{L}d_{\alpha}\label{11}
\end{eqnarray}
where
\begin{eqnarray}
C_{i\alpha}&=&(U^\dag D^\dag V)_{i\alpha}=((DU)^\dag V)_{i\alpha}\notag\\
C^{*}_{i\alpha}&=&(C^\dag)_{\alpha i}\label{12}
\end{eqnarray}
Similarly
\begin{eqnarray}
J^{X\mu}(\nu)=(C_{\nu})_{i\alpha}\bar\nu_i\Gamma^{\mu}_{L}u_\alpha\label{13}
\end{eqnarray}
where
\begin{eqnarray}
(C_{\nu})_{i\alpha}=(DU)^\dag_{i\alpha}=(U^\dag
D^\dag)_{i\alpha}\label{14}
\end{eqnarray}
Hence the effective Hamiltonian involving the current $J^{X\mu}(e)$
is given by
\begin{eqnarray}
\mathcal{H}=\frac{g^2_X}{gm^2_X}[C_{i\alpha}\bar
e_i\Gamma^{\mu}_{L}d_\alpha][\bar
d_{\beta}\Gamma_{L\mu}e_j]C^{*}_{j\beta}\label{15}
\end{eqnarray}
After Fierz reshuffling
\begin{eqnarray}
\mathcal{H}=\frac{G_X}{\sqrt2}C_{i\alpha}C^{*}_{j\beta}[\bar
d_{\beta}\Gamma^{\mu}_{L}d_\alpha][\bar
e_i\Gamma_{L\mu}e_j]\label{16}
\end{eqnarray}
where
\begin{eqnarray}
\frac{G_X}{\sqrt2}&=&\frac{g^2_X}{gm^2_X}\notag\\
G_X&=&\left(\frac{g_X}{g}\right)^2\frac{m^2_W}{m^2_X}G_F\label{17}
\end{eqnarray}
Similarly the effective Hamiltonian involving $J^{X\mu}(\nu)$ is
given by
\begin{eqnarray}
\mathcal{H}=\frac{G_X}{\sqrt2}(C_\nu)_{i\alpha}C^{*}_{j\beta}[\bar
d_{\beta}\Gamma^{\mu}_{L}u_\alpha][\bar \nu_i\Gamma_{L\mu}e_j]\notag
\end{eqnarray}

The effective Hamiltonian given in Eq.(\ref{16}) is the basic
Hamiltonian for \text{FCNC} induced processes. The most stringent
experimental limits on the \text{LF} violating decays are for the
$K$ mesons decays\cite{[2]}:
\begin{subequations}
\begin{eqnarray}
\text{B.R}\quad(K^{0}_{L}\rightarrow e^{\mp}\mu^{\pm})&< &4.7\times
10^{-12}\label{18-a}\\
\text{B.R}\quad(K^{0}_{L}\rightarrow \mu^{-}\mu^{+})&= &(6.84\pm
0.11)\times 10^{-9}\label{18-b}
\end{eqnarray}
\end{subequations}
\begin{subequations}
\begin{eqnarray}
\text{B.R}\quad(K^{-}\rightarrow \pi^{-}\mu^{-} e^{+})< 1.3\times
10^{-11}\label{19-a}\\
\text{B.R}\quad(K^{-}\rightarrow \pi^{-}e^{-} \mu^{+})< 1.5\times
10^{-10}\label{19-b}
\end{eqnarray}
\end{subequations}
For $K$ mesons, the effective Hamiltonian is given by
\begin{align}
\mathcal{H}={G_X\over\sqrt2}\left\{[\bar
d\Gamma_{L}^{\mu}s][(C_{es}C^{*}_{ed})(\bar
e\Gamma_{L\mu}e)+(C_{es}C^{*}_{\mu d})(\bar e\Gamma_{L\mu}\mu)
+(C_{\mu s}C^{*}_{ed})(\bar\mu\Gamma_{L\mu}e)+(C_{\mu s}C^{*}_{\mu
d})(\bar\mu\Gamma_{L\mu}\mu)]\right\}\label{20}
\end{align}

\begin{figure}[htb!]
\centering
\includegraphics[scale=0.5]{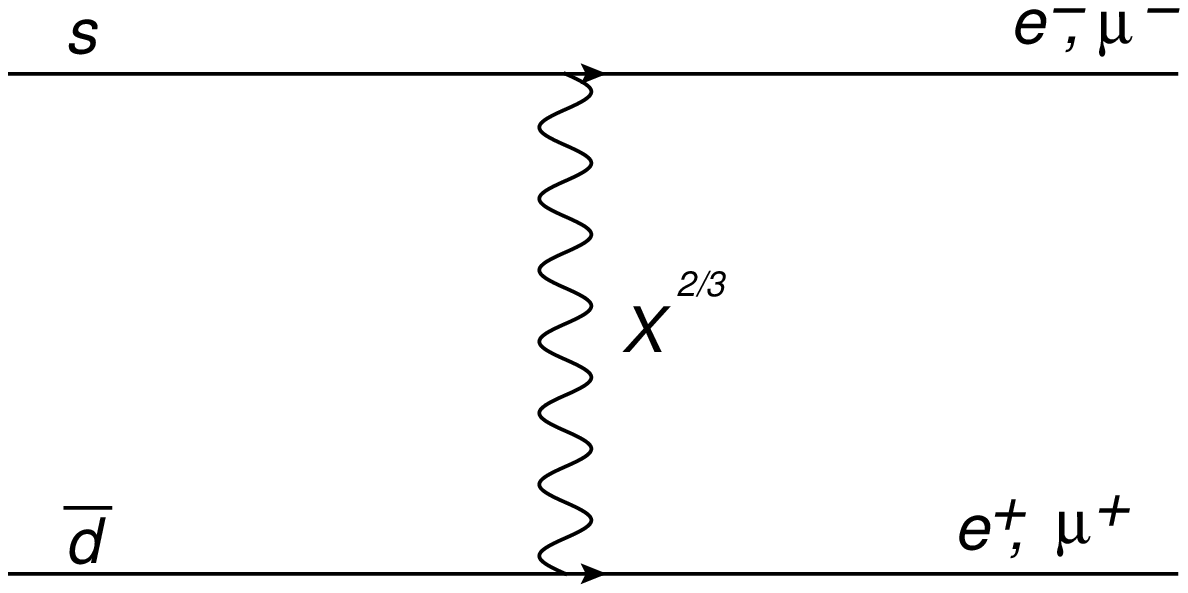}
\caption{} \label{fig:1}
\end{figure}

\begin{figure}[htb!]
\centering
\includegraphics[scale=0.5]{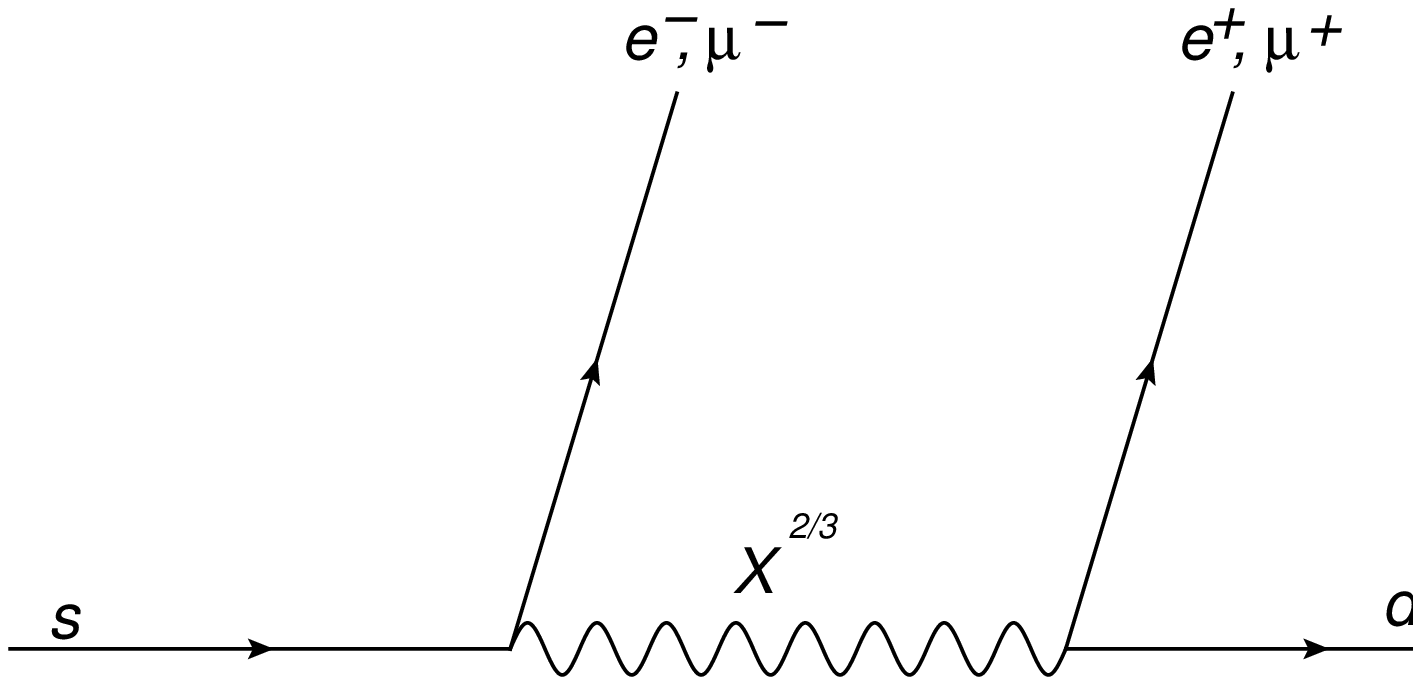}
\caption{} \label{fig:2}
\end{figure}

The above Hamiltonian is relevant for the decays

\begin{eqnarray}
K^{0}_{L}&\rightarrow& e^-e^+\notag\\
&\rightarrow& \mu^-\mu^+\notag\\
&\rightarrow& e^-\mu^+ , \mu^-e^+\notag
\end{eqnarray}
and
\begin{eqnarray}
K^-&\rightarrow& \pi^- e^-e^+\notag\\
&\rightarrow& \pi^- e^-\mu^+ (\mu^-e^+)\notag\\
&\rightarrow& \pi^- \mu^-\mu^+\notag
\end{eqnarray}
shown in Figs (\ref{fig:1}) and (\ref{fig:2}).

From the Hamiltonian (\ref{20}) , we obtain the following branching
ratios
\begin{eqnarray}
\text{B.R}\quad(K^{0}_{L}\rightarrow e^{\pm}\mu^{\mp})&\approx &
\frac{\tau_{K^0_L}}{\hbar}\left({G_X\over G_F}\right)^2\left(1-{m^2
_{\mu}\over m^2_K}\right)^2{1\over 2}\Big[{G^2_F\over 8\pi}f^2_K
m^2_{\mu}m_K\Big]\times |C_{\mu s}C^*_{ed}+C_{es}C_{\mu
d}|^2\label{21}\\
\text{B.R}\quad(K^{0}_{L}\rightarrow \mu^-\mu^+)&\approx &
\frac{\tau_{K^0_L}}{\hbar}\left({G_X\over G_F}\right)^2\left(1-2{m^2
_{\mu}\over m^2_K}\right)2\Big[{G^2_F\over 8\pi}f^2_K 2
m^2_{\mu}m_K\Big]\times |C_{\mu s}C^*_{\mu d}|^2\label{22}\\
\text{B.R}\quad(K^-\rightarrow \pi^- \mu^-e^+)&\approx &
\left({G_X\over G_F}\right)^2\Big|\frac{C_{\mu s}C^*_{ed}}{V_{\mu
s}}\Big|^2\times \text{B.R.}(K^-\rightarrow
\pi^0\mu^-\bar\nu_{\mu})\label{23}\\
\text{B.R}\quad(K^-\rightarrow \pi^- e^-\mu^+)&\approx &
\left({G_X\over G_F}\right)^2\Big|\frac{C_{e s}C^*_{\mu d}}{V_{\mu
s}}\Big|^2\times \text{B.R.}(K^-\rightarrow
\pi^0\mu^-\bar\nu_{\mu})\label{24}
\end{eqnarray}
In the above expressions, we have neglected $m^2_e$ as compared to
$m^2_\mu$.

Using the following values:
\begin{eqnarray}
\tau_{K^0_L}=5.116\times 10^{-8}s, G_F=1.166\times 10^{-5}
\text{GeV}^{-2}\notag\\
f_K=0.160\text{GeV}, m_K=0.497\text{GeV},
m_{\mu}=0.106\text{GeV}\notag
\end{eqnarray}
we obtain from Eqs.(\ref{21}) and (\ref{22}):
\begin{eqnarray}
\text{B.R}\quad(K^{0}_{L}\rightarrow e^{\pm}\mu^{\mp})&\approx &
27\left({G_X\over G_F}\right)^2 |C_{\mu s}C^*_{ed}+C_{es}C^*_{\mu
d}|^2\label{25}\\
\text{B.R}\quad(K^{0}_{L}\rightarrow \mu^-\mu^+)&\approx &
216\left({G_X\over G_F}\right)^2 |C_{\mu s}C^*_{\mu d}|^2\label{26}
\end{eqnarray}

The mulitplets $(2,\bar2)$ follows the patterns:
\begin{eqnarray*}
D(e):\left(\begin{array}{cc}u&\nu_{e}\\
d^{\prime}&e\end{array}\right),& \left(\begin{array}{cc}c&\nu_{\mu}\\
s^{\prime}&\mu\end{array}\right),& \left(\begin{array}{cc}t&\nu_{\tau}\\
b^{\prime}&\tau\end{array}\right)\\
D(12):\left(\begin{array}{cc}u&\nu_{\mu}\\
d^{\prime}&\mu\end{array}\right),& \left(\begin{array}{cc}c&\nu_{e}\\
s^{\prime}&e\end{array}\right),& \left(\begin{array}{cc}t&\nu_{\tau}\\
b^{\prime}&\tau\end{array}\right)\\
D(23):\left(\begin{array}{cc}u&\nu_{e}\\
d^{\prime}&e\end{array}\right),& \left(\begin{array}{cc}c&\nu_{\tau}\\
s^{\prime}&\tau\end{array}\right),& \left(\begin{array}{cc}t&\nu_{\mu}\\
b^{\prime}&\mu\end{array}\right)\\
D(231):\left(\begin{array}{cc}u&\nu_{\mu}\\
d^{\prime}&\mu\end{array}\right),& \left(\begin{array}{cc}c&\nu_{\tau}\\
s^{\prime}&\tau\end{array}\right),& \left(\begin{array}{cc}t&\nu_{e}\\
b^{\prime}&e\end{array}\right)\\
\end{eqnarray*}
The matrix $U=U_{PMNS}$, involves the parameters $(c_{12}, s_{12})$,
$(c_{23}, s_{23})$, $(c_{13},
s_{13}e^{i\delta})$\cite{[2],[3]},where
\begin{eqnarray}
c_{12}&=&\cos\theta_{12}\approx 0.835, s_{12}=\sin\theta_{12}\approx
0.551,\notag\\
c_{23}&=&\cos\theta_{23}\approx 0.707, s_{23}=\sin\theta_{23}\approx
0.707,\notag\\
c_{13}&=&\cos\theta_{13}, s_{13}=\sin\theta_{13}\notag
\end{eqnarray}
We take $\theta_{13}\approx 0$, although a recent experiment
\cite{[4]} gives $\theta_{13}\approx 9^\circ$. With these values
\begin{eqnarray}
U_{PMNS}=\left(\begin{array}{ccc}0.835&0.551&0\\
-0.389&0.590&0.707\\
0.389&-0.590&0.707\end{array}\right)\label{27}
\end{eqnarray}
For the \text{CKM} matrix $V$, we use the following values
\cite{[2]}
\begin{eqnarray}
V=\left(\begin{array}{ccc}0.974&0.225&3.89\times
10^{-3}e^{-i\gamma}\\
-0.223&1&4.06\times 10^{-2}\\
8.4\times 10^{-3}e^{i\alpha}&-3.87\times
10^{-2}&1\end{array}\right)\label{28}
\end{eqnarray}

In this paper, the coefficients $C_{i\alpha}$ are obtained for the
following two cases I and II\\
\textbf{Case I}: $D(e)$, $D(23)$,$D(12)$,$D(231)$ (i) $U=1$ (ii)
$U=U_{PMNS}$. Using these coefficients and Eq.(\ref{25}), upper
limit on $\left(G_X/G_F\right)$ is obtained from the inequality
(\ref{18-a}).Then using Eqs. (\ref{26}), (\ref{23}) and (\ref{24}),
we get upper limits on the branching ratios $K^{0}_{L}\to
\mu^{-}\mu^{+}$, $K^{-}\to
\pi^{-}\mu^{-}e^{+}(\pi^{-}e^{-}\mu^{+})$. The results are given in
Table1.\\
\textbf{Case II}: $D=1$
\begin{eqnarray}
(i) U=\left(\begin{array}{ccc} c_{12}&-s_{12}&0\\
s_{12}&c_{12}&0\\
0&0&1\end{array}\right),&& (ii) U=\left(\begin{array}{ccc} 1&0&0\\
0&c_{23}&-s_{23}\\
0&s_{23}&c_{23}\end{array}\right)\label{29}
\end{eqnarray}
The results are given in Table II\\

\begin{table}\caption{The upper bounds on $\frac{G_{X}}{G_{F}}$ and upper limits on the B.R. ($K^0_L\to\mu^+\mu^-$); B.R. ($K^-\pi^-\mu^-e^+(\pi^-e^-\mu^+)$) for case I: $D=D(e), D(23), D(12), D(231);$ $U=1$, $U=U_{PMNS}$ }
\scalebox{0.75}{
\begin{tabular}{|c|c|c|c|c|c|c|c|}
\hline\hline \multicolumn{3}{|c|}{\multirow{2}*{}}&
$\frac{G_{X}}{G_{F}}$ $<$ & $(\frac{g}{g_{X}})(\frac{m_{X}}{m_{W}})$
$>$ & $B.R(K_{L}^{0}\rightarrow \mu ^{+}\mu ^{-})$ &
$B.R(K^{-}\rightarrow \pi ^{-}\mu ^{-}e^{+})$ &
$B.R(K^{-}\rightarrow \pi^{-}e^{-}\mu ^{+})$\\
\multicolumn{3}{|c|}{}& Upper bound & lower bound & $<$ & $<$ & $<$ \\
 \hline\hline

\multirow{3}{*}{(i)} &  & $U=1$ & $4.91\times 10^{-7}$ & $1.43\times
10^{3}$ & $2.60\times 10^{-12}$ & $1.59\times 10^{-13}$
& $4.23\times 10^{-16}$ \\
&$D(e)$ & & & & & & \\
& & $U_{PMNS}$ & $7.75\times 10^{-7}$ & $1.13\times 10^{3}$ &
$9.82\times 10^{-12}$ & $1.63\times 10^{-13}$ & $3.09\times
10^{-11}$ \\

\hline\hline \multirow{3}{*}{(ii)} &  & $U=1$ & $1.05\times 10^{-5}$
& $3.09\times 10^{2}$ & $2.54\times 10^{-15}$ & $1.09\times
10^{-13}$
& $3.69\times 10^{-16}$ \\
&$D(23)$ & & & & & & \\
& & $U_{PMNS}$ & $9.07\times 10^{-6}$ & $3.32\times 10^{2}$ & $1.95\times 10^{-9}$
& $7.05\times 10^{-12}$ & $9.03\times 10^{-12}$ \\
\hline\hline

\multirow{3}{*}{(iii)} &  & $U=1$ & $4.91\times 10^{-7}$ &
$1.43\times
10^{3}$ & $2.60\times 10^{-12}$ & $4.23\times 10^{-16}$ & $1.59\times 10^{-13}$ \\
&$D(12)$ & & & & & & \\
& & $U_{PMNS}$ & $6.13\times 10^{-6}$ & $4.04\times 10^{2}$ & $8.28\times 10^{-9}$
& $4.18\times 10^{-12}$ & $2.88\times 10^{-12}$ \\
\hline\hline

 \multirow{3}{*}{(iv)} &  & $U=1$ & $1.10\times 10^{-5}$ &
$3.01\times 10^{2}$ & $1.26\times 10^{-9}$ & $3.04\times 10^{-16}$
& $1.22\times 10^{-13}$ \\
&$D(231)$ & & & & & & \\
& & $U_{PMNS}$ & $1.00\times 10^{-6}$ & $10^{3}$ & $2.5\times
10^{-11}$ & $3.51\times 10^{-14}$ & $2.61\times
10^{-14}$ \\

\hline\hline

\end{tabular}}
\end{table}
\begin{table}\caption{The upper limit for the branching ratios $K^0_L\to\mu^-\mu^+$, $K^-\to\pi^0\mu^-e^+(\pi^0e^-\mu^+)$ decays for case II.}
\scalebox{0.75}{
\begin{tabular}[t]{|c|c|c|c|c|c|}
\hline\hline \multirow{2}{*}{Lepton Mixing Matrix U} &
$\frac{G_{X}}{G_{F}}$ $<$ & $(\frac{g}{g_{X}})(\frac{m_{X}}{m_{W}})$
$>$ & $B.R(K_{L}^{0}\rightarrow \mu ^{+}\mu ^{-})$ &
$B.R(K^{-}\rightarrow \pi ^{0}\mu ^{-}e^{+})$ &
$B.R(K^{-}\rightarrow \pi^{0}e^{-}\mu ^{+})$ \\
& Upper bound & lower bound & $<$ & $<$ & $<$ \\
\hline \hline
 (i) & $9.5\times 10^{-6}$ & $3.2\times 10^{2}$ &
$5.1\times 10^{-9}$ & $1.5\times 10^{-11}$  & $1.8\times 10^{-11}$ \\
\hline\hline (ii) & $6.7\times 10^{-7} $ & $1.2\times 10^{3}$ &
$1.1\times 10^{-12}$
& $1.3\times 10^{-13}$ & $3.9\times 10^{-16}$ \\
\hline\hline
\end{tabular}}
\end{table}

From Tables I and II, it is clear that for the case I, $D=D(23)$,
$D(12)$, $U=U_{PMNS}$ and for the case II, $D=1$, $U=U(i)$, the
upper bound on ($\frac{G_X}{G_F}$) is the lowest and the branching
ratios are compatible with the experimental bounds except for the
B.R. ($K_L^0\to\mu^+\mu^-$) which exceeds the experimental value for
case I: $D(12)$, $U=U_{PMNS}$

Finally from the effective Hamiltonians (3.6) and (3.7), we obtain
the branching ratios for L.F.violating $\tau$ decays and for
$\pi^+\to\mu^+\nu_e(\nu_\tau)$, $n\to
pe^-\bar{\nu}_\mu(\bar{\nu}_\tau)$:
\begin{eqnarray}
B.R.(\tau^-\to\bar{K}^{*0}l^-)&=&(\frac{G_X}{G_F})^2|\frac{C_{\tau
s}C_{ld}^*}{V_{us}}|^2B.R(\tau^-\to K^{-*}\bar{\nu}_\tau) \\
B.R.(\tau^-\to\rho^0l^-)&=&(\frac{G_X}{G_F})^2|\frac{C_{\tau
d}C_{ld}^*}{V_{ud}}|^2B.R(\tau^-\to \rho^-\nu_\tau) \\
B.R.(\pi^+\to\mu^+\nu_{e,\tau})&=&(\frac{G_X}{G_F})^2|\frac{C_{\nu
u}C_{\mu d}^*}{V_{ud}}|^2B.R(\pi^+\to\mu^+\nu_\mu)
\\
B.R.(n\to
pe^-\bar{\nu}_{\mu,\tau})&=&(\frac{G_X}{G_F})^2|\frac{C_{ed}C_{\nu
u}^*}{V_{ud}}|^2B.R(n\to pe^-\bar{\nu}_e)
\\
B.R.(\tau^-\to
K^{*-}\nu_{e,\mu})&=&(\frac{G_X}{G_F})^2|\frac{C_{\tau
s}C_{\nu u}^*}{V_{us}}|^2B.R(\tau^-\to K^{*-}\nu_\tau) \\
B.R.(\tau^-\to\rho^-\nu_{e,\mu})&=&(\frac{G_X}{G_F})^2|\frac{C_{\tau
d}C_{\nu u}^*}{V_{ud}}|^2B.R(\tau^-\to\rho^-\nu_\tau)
\end{eqnarray}
where $l=e,\mu$. In Eqs. (3.20-3.25), we have neglected the terms of
order $m_l^2/m_\tau^2$. From these equations, we get the following
values for the branching ratios:

\begin{tabular}{c c c}
case I, $U_{PMNS}$ & (ii) D(23) & (iii) D(12)  \\
B.R.($\tau^-\to\bar{K}^{*0}e^-(\mu^-)$) & $<4.4\times10^{-12}$ & B.R.($\tau^-\to K^{*0}e^-(\mu^-)<2.1\times10^-12$) \\
B.R.($\tau^-\to\rho^0e^-{\mu^-}$) & $<2.7\times10^{-13}$ & $<1.4(0.92)\times10^{-12}$ \\
B.R.($\pi^+\to\mu^+\nu_{e,\tau}$) & $<2.9\times10^{-11}(6.4\times10^{-12})$ & $<1.1\times10^{-12}$ \\
B.R.($\tau^-\to K^{*-}\nu{e,\mu}$) & $<6.6\times10^{-12}$ & $<5.6(1.1)\times10^{-12}$ \\
B.R.($\tau^-\to\rho^-\nu_{e,\mu}$) & $<7.9(1.3)\times10^{-12}$ &
B.R.($\tau^-\to\rho^-\nu_{e,\mu}$)($<0.7(3.1)\times10^{-12}$)
\end{tabular}
\section{Lepton Flavor (LF) violating B decays}

The effective  Hamiltonian for \text{LF} violating $B-\text{decays}$
is given by
\begin{eqnarray}
\mathcal{H}_{eff}={G_X\over\sqrt2}(C_{ib}C^*_{jd})[(\bar
d\Gamma^\mu_L b)(\bar e_i\Gamma_{L\mu}e_j)]+(C_{ib}C^*_{js})[(\bar
s\Gamma^\mu_L b)(\bar e_i\Gamma_{L\mu}e_j)]\label{29}
\end{eqnarray}
In this paper , only the decays $B^\circ\rightarrow\ell_j^-\ell_i^+$
$(\bar B^\circ\rightarrow\ell_i^-\ell_j^+)$ are analysed.

The branching ratio is given by
\begin{align}
\text{B.R.}(B^\circ\rightarrow\ell_j^-\ell_i^+)&={\tau_{B^\circ}\over\hbar}\left({G_X\over
G_F}\right)^2\left(1-{m^2_i+m^2_j\over
m^2_{B^\circ}}\right)\left(1-{m^2_i-m^2_j\over
m^2_{B^\circ}}\right)&\Big[{G^2_F\over
8\pi}m_{B^\circ}f^2_B(m^2_i+m^2_j)\times|g_{ji}|^2\Big]\label{30}
\end{align}
where
\begin{eqnarray}
g_{ji}=C_{j\beta}C^*_{ib}\quad
g_{ij}=C_{ib}C^*_{j\beta}=g^*_{ji},\quad \beta=\text{d or s} \notag
\end{eqnarray}
Since branching ratio is proportional to $m^2_l$, we will take
$i=\tau$, $j=e,\mu,\tau$; $g_{ji}=g_{l\tau}$, $l=e,\mu,\tau$. To
obtain the branching ratios, we use the following values:
\begin{eqnarray}
\tau_{B^\circ_d}=1.52\times 10^{-12}s,\quad m_{B^\circ_d}=5.279
\text{Gev},\quad f_{B_d}=0.220\text{Gev}\notag\\
\tau_{B^\circ_s}=1.47\times 10^{-12}s,\quad m_{B^\circ_s}=5.366
\text{Gev},\quad f_{B_s}=0.234\text{Gev}\notag
\end{eqnarray}
Using above values, we get from Eq.(\ref{30})
\begin{eqnarray}
\text{B.R.}[B^\circ\rightarrow\ell^-\tau^+(\tau^-\ell^+)]=\left({G_X\over
G_F}\right)^2
F\Big[|g_{\ell\tau}|^2\Big(|g_{\tau\ell}|^2\Big)\Big]\label{31}
\end{eqnarray}
where
\begin{eqnarray}
F&=&7.87,\quad B^\circ_d\notag\\
&=&8.87, \quad B^\circ_s \label{32}
\end{eqnarray}
For $B^\circ\rightarrow \tau^-\tau^+$, $\mu^-\mu^+$
\begin{eqnarray}
F&=&15.48,\quad 0.071\quad B^\circ_d\notag\\
&=&17.43,\quad 0.079\quad B^\circ_s \label{33}
\end{eqnarray}
Now
\begin{eqnarray}
\left({G_X\over G_F}\right)^2<(5.71\times 10^{-13}),\quad
(8.23\times 10^{-11}),\quad (3.76\times 10^{-11})\label{34}
\end{eqnarray}
for $D(e)$, $D(23)$, $D(12)$, $U=U_{PMNS}$ respectively. The factors
$g_{\ell\tau}$ $(g_{\tau\ell})$ in Eq.(\ref{31}) for the three cases
(i), (ii), (iii) are given below
\\
(i)\quad $B^\circ_d$:
\begin{align}
g_{e\tau}(g_{\tau e})&\approx0.602 (-0.051),\quad
g_{\mu\tau}(g_{\tau\mu})\approx 0.259 (0.077)\notag\\
g_{\tau^-\tau^+}&\approx -0.101,\quad\quad\quad\quad\quad
g_{\mu^-\mu^+}\approx -0.197\notag
\end{align}
\quad\quad $B^\circ_s$:
\begin{align}
g_{e\tau}(g_{\tau e})&\approx-0.124 (0.245),\quad
g_{\mu\tau}(g_{\tau\mu})\approx 0.495 (-0.372)\notag\\
g_{\tau^-\tau^+}&\approx 0.488,\quad\quad\quad\quad\quad\quad
g_{\mu^-\mu^+}\approx -0.350\notag
\end{align}
(ii)\quad $B^\circ_d$:
\begin{align}
g_{e\tau}(g_{\tau e})&\approx0.520 (0.058),\quad
g_{\mu\tau}(g_{\tau\mu})\approx 0.492 (-0.089)\notag\\
g_{\tau^-\tau^+}&\approx -0.115,\quad\quad\quad\quad
g_{\mu^-\mu^+}\approx 0.379\notag
\end{align}
\quad\quad $B^\circ_s$:
\begin{align}
g_{e\tau}(g_{\tau e})&\approx0.436 (-0.245),\quad
g_{\mu\tau}(g_{\tau\mu})\approx -0.365 (0.385)\notag\\
g_{\tau^-\tau^+}&\approx 0.500,\quad\quad\quad\quad\quad
g_{\mu^-\mu^+}\approx 0.079\notag
\end{align}
(iii)\quad $B^\circ_d$:
\begin{align}
g_{e\tau}(g_{\tau e})&\approx-0.408 (0.291),\quad
g_{\mu\tau}(g_{\tau\mu})\approx 0.327 (-0.389)\notag\\
g_{\tau^-\tau^+}&\approx 0.498,\quad\quad\quad\quad\quad\quad
g_{\mu^-\mu^+}\approx -0.257\notag
\end{align}
\quad\quad $B^\circ_s$:
\begin{align}
g_{e\tau}(g_{\tau e})&\approx 0.530 (0.056),\quad
g_{\mu\tau}(g_{\tau\mu})\approx 0.511 (-0.075)\notag\\
g_{\tau^-\tau^+}&\approx 0.095,\quad\quad\quad\quad\quad
g_{\mu^-\mu^+}\approx 0.420\notag
\end{align}
With above values for $g_{\ell i}g_{i\ell}$, the branching ratios
can be calculated. The upper limit on the branching ratios for
various decay channels are given in Table III.
\begin{table}
\centering\caption{Upper limit on B.R.$(B_{d,s}^0\to
l^-\tau^+(\tau^-l^+))$} case I: for $D=D(e)$, $D(23)$, $D(12)$,
$U_{PMNS}$
\begin{tabular}{c c c c}
Case & (i) & (ii) & (iii) \\ [0.9ex]
$B^\circ_{d,s}\rightarrow e^-\tau^+$ & $1.60\times 10^{-12}$ & $1.75\times 10^{-10}$&$ 4.92\times 10^{-11}$ \\
 & $\left(7.78\times 10^{-14}\right)$ & $\left(1.39\times 10^{-10}\right)$ & $\left(9.37\times 10^{-11}\right)$ \\
 \\
$\quad\quad\rightarrow \tau^- e^+$ & $1.17\times 10^{-14}$ & $2.18\times 10^{-12}$&$ 2.50\times 10^{-11}$ \\
 & $\left(3.04\times 10^{-13}\right)$ & $\left(4.71\times 10^{-11}\right)$ & $\left(1.04\times 10^{-12}\right)$ \\
 \\
$\quad\quad\rightarrow \mu^-\tau^+$ & $3.01\times 10^{-13}$ & $1.57\times 10^{-10}$&$ 3.16\times 10^{-11}$ \\
 & $\left(1.17\times 10^{-12}\right)$ & $\left(9.72\times 10^{-11}\right)$ & $\left(8.71\times 10^{-11}\right)$ \\
 \\
$\quad\quad\rightarrow \tau^-\mu^+$ & $2.66\times 10^{-14}$ & $5.13\times 10^{-12}$&$ 4.48\times 10^{-11}$ \\
 & $\left(7.01\times 10^{-13}\right)$ & $\left(1.08\times 10^{-10}\right)$ & $\left(1.88\times 10^{-12}\right)$ \\
 \\
 $\quad\quad\rightarrow \tau^-\tau^+$ & $9.02\times 10^{-14}$ & $1.68\times 10^{-11}$&$ 1.44\times 10^{-10}$ \\
 & $\left(2.37\times 10^{-12}\right)$ & $\left(3.59\times 10^{-10}\right)$ & $\left(5.91\times 10^{-12}\right)$ \\
 \\
 $\quad\quad\rightarrow \mu^-\mu^+$ & $1.57\times 10^{-15}$ & $8.39\times 10^{-13}$&$ 1.76\times 10^{-13}$ \\
 & $\left(5.52\times 10^{-15}\right)$ & $\left(4.08\times 10^{-14}\right)$ & $\left(4.87\times 10^{-13}\right)$ \\
\end{tabular}
\end{table}
The experimental limits on L.F. violating $B^\circ\rightarrow
\ell^-_j\ell^+_i (j\neq i)$ decays are not stringent as for the
$K-$decays. For $B^\circ$ decays, experimental limits are
\cite{[2],[5],[7]}
\begin{eqnarray}
\text{B.R.}(B^\circ_d&\rightarrow &e^{\mp}\tau^{\pm})\quad<2.8\times
10^{-5}\notag\\
\text{B.R.}(B^\circ_d&\rightarrow
&\mu^{\mp}\tau^{\pm})\quad<2.2\times
10^{-5}\notag\\
\text{B.R.}(B^\circ_d&\rightarrow &\tau^{-}\tau^{+})\quad\quad ?\notag\\
\text{B.R.}(B^\circ_s&\rightarrow
&\mu^{-}\mu^{+})_{\text{exp}}\quad=(3.2^{+1.5}_{-1.2})\times10^{-9}\notag\\
\text{B.R.}(B^\circ_s&\rightarrow
&\mu^{-}\mu^{+})_{\text{SM}}\quad=(3.32\pm 0.17)\times
10^{-9}\notag\\
\text{B.R.}(B^\circ_d&\rightarrow
&\mu^{-}\mu^{-})_{\text{exp}}\quad<0.8\times
10^{-9}\notag\\
\text{B.R.}(B^\circ_d&\rightarrow
&\mu^{-}\mu^{-})_{\text{SM}}\quad=(0.10\pm 0.01)\times 10^{-9}\notag
\end{eqnarray}
It is clear from the above Table III, the upper limits on branching
ratios for $B^\circ_{s,d}\rightarrow \mu^-\mu^+$ are much below than
the experimental upper limits for these decays.

However, a distinctive feature of this model is violation of charge
symmetry for the decays $B^\circ_d\rightarrow e^-(\mu^-)\tau^+$ and
$B^\circ_d\rightarrow \tau^- e^+(\mu^+)$ and  $B^\circ_s\rightarrow
e^-(\mu^-)\tau^+$ and $B^\circ_s\rightarrow \tau^- e^+(\mu^+)$ for
(ii) and (iii) respectively. The branching ratios are significantly
different for the channels $e^-(\mu^-)\tau^+$ and $\tau^-
e^+(\mu^+)$; the branching ratios are much smaller for the channels
$\tau^- e^+(\mu^+)$ than for $e^-(\mu^-)\tau^+$.

Finally for the case II, $D=1$, $U=U(i)$;
$(G_X/G_F)<9.5\times10^{-6}$. For this case:
\begin{eqnarray}
B_d^0:g_{e\tau}(g_{\tau
e})\approx0.69(0);\quad g_{\mu\tau}(g_{\tau\mu})\approx0.70(0)\notag\\
B_s^0:g_{e\tau}(g_{\tau e})\approx0.74(0);\quad
g_{\mu\tau}(g_{\tau\mu})\approx0.77(0)\notag
\end{eqnarray}

The upper limitations on the branching ratios for the case II are
\begin{eqnarray}
B.R. (B_d^0\to
e^-\tau^+(\mu^-\tau^+))<3.4\times10^{-10}(3.7\times10^{-10})\notag\\
B.R. (B_s^0\to
e^-\tau^+(\mu^-\tau^+))<4.4\times10^{-10}(4.7\times10^{-10})
\end{eqnarray}

The branching ratios for $B_{d,s}^0\tau^-e^+(\tau^-\mu^+)\approx0$.
Again for this case charge symmetry is badly broken.

The time integrated decay rates due to quantum interference of
$B^\circ$ and $\bar B^\circ$ for the LF violating decays of
$B^\circ$ to lepton pairs are a promising area to test the model
based on the extended electroweak unification group $SU_L(2)\times
U_{Y_1}\times SU_X(2)$.

Now \cite{[3]}
\begin{eqnarray}
\Gamma_f(t)(\bar\Gamma_{\bar f}(t))&=&{1\over2}e^{-\Gamma
t}|F|^2\left(|g_{\ell\tau}|^2+|g_{\tau\ell}|^2\right)+\left(|g_{\ell\tau}|^2-|g_{\tau\ell}|^2\right)\cos\Delta mt\notag\\
&\pm & i\sin\Delta mt
\left(e^{2i\beta}g_{\ell\tau}g_{\tau\ell}-e^{-2i\beta}g^*_{\ell\tau}g^*_{\tau\ell}\right)\notag
\end{eqnarray}
\begin{eqnarray}
\Gamma_{\bar f}(t)(\bar\Gamma_{f}(t))&=&{1\over2}e^{-\Gamma
t}|F|^2\left(|g_{\ell\tau}|^2+|g_{\tau\ell}|^2\right)-\left(|g_{\ell\tau}|^2-|g_{\tau\ell}|^2\right)\cos\Delta mt\notag\\
&\pm & i\sin\Delta mt
\left(e^{2i\beta}g_{\ell\tau}g_{\tau\ell}-e^{-2i\beta}g^*_{\ell\tau}g^*_{\tau\ell}\right)\notag
\end{eqnarray}
where $f=\ell^-\tau^+$, $\bar f=\tau^-\ell^+$, $\ell=e,\mu$

First we note that $\beta=0$, for $B^\circ_s$, so $\sin\Delta mt$
term vanishes for $B^\circ_s$. For the decays for which
$g_{\tau\ell}\ll g_{\ell\tau}$, we have \cite{[3]}
\begin{eqnarray}
\Gamma_{f}(t)&=&{1\over2}e^{-\Gamma
t}|F|^2|g_{\ell\tau}|^2(1+\cos\Delta mt)\notag\\
\Gamma_{\bar f}(t)&=&{1\over2}e^{-\Gamma
t}|F|^2|g_{\ell\tau}|^2(1-\cos\Delta mt)\notag
\end{eqnarray}
This is the case for $B^\circ_{d,s}\rightarrow \ell^-\tau^+$
$(\tau^-\ell^+)$ in (ii) and (iii) respectively. For the time
integrated decay rates, we have
\begin{eqnarray}
\frac{N(\tau^-\ell^+)_{d,s}}{N(\ell^-\tau^+)_{d,s}}=\frac{\int_0^\infty
\Gamma_{\bar f}(t)dt}{\int_0^\infty \Gamma_{f}(t)dt}&=&\frac{(\Delta
m/\Gamma)^2_{d,s}}{[2+(\Delta m/\Gamma)^2_{d,s}]}\notag\\
&=&\frac{x^2_{d,s}}{2+x^2_{d,s}}\notag\\
&\approx&0.23(1)\notag
\end{eqnarray}
where we have used the experimental values $x_d\approx 0.77$,
$x_s\approx 26$ for $B^\circ_d$ and $B^\circ_s$ respectively.

Hence for time integrated decay rates for $B^\circ_d$, the
$\tau^-/\tau^+=e^-/e^+\approx 0.23$ i.e. depletion of $\tau^-$ or
$e^+$, compared to $\tau^+$ or $e^-$ i.e. charge symmetry is badly
violated for the time integrated decay rates $B^\circ_d\rightarrow
\tau^-\ell^+$ $(\ell^-\tau^+)$.

However, for time integrated decay rates $B^\circ_s\rightarrow
\tau^-\ell^+$ $(\ell^-\tau^+)$, there is only extremely small
deviation from 1, although $\text{B.R.}(B^\circ_s\rightarrow
\tau^-\ell^+/\text{B.R.}(B^\circ_s\rightarrow\ell^-\tau^+\ll 1$.

\section{Left-Right symmetric group $SU_L(2)\times SU_R(2)\times U_{Y_{1}}(1)\times
SU_X(2)$}

The left-handed and right-handed fermions are assigned to following
representation of the group:
\begin{eqnarray}
\left(\begin{array}{cc}u_i & \nu_i\\
d^{\prime}_i & e_i\end{array}\right)_L: \left(2, 1, \bar
2\right)\notag\\
\left(\begin{array}{cc}u_i & N_i\\
d^{\prime}_i & e_i\end{array}\right)_R: \left(1, 2, \bar
2\right)\notag
\end{eqnarray}
\begin{eqnarray}
Q&=&{1\over 2}(\tau_{L_3}+\tau_{R_3}+Y_1+\tau_{X_3})\notag
\end{eqnarray}
\begin{eqnarray}
Y_1=\begin{array}{rcl} 0 & \mbox{for leptons}\\ -1/3 & \mbox{for
quarks}\notag
\end{array}
\end{eqnarray}
The interaction Lagrangian is given by
\begin{align}
L_{\text{int}}=&-g\sin\theta_W J^\mu_{em}A_{\mu}-\frac{g}{2\sqrt 2}
\Big[\Big(\bar\nu_i\Gamma^{\mu}_Le_i+\bar u_i\Gamma^{\mu}_L
d_i^{\prime}\Big)W_{L\mu}^{+}+\Big(\bar N_i\Gamma^{\mu}_Re_i+\bar
u_i\Gamma^{\mu}_R d_i^{\prime}\Big)W_{R\mu}^{+}+h.c\Big]\notag\\
&-{g\over\cos\theta_W}J^{Z\mu}Z_{\mu}-\frac{g}{\sqrt{1-\tan^2\theta_w}}
J^{Z^{\prime}\mu}Z^{\prime}_{\mu}-{g_X\over
g}\frac{g}{\sqrt{1-g^{\prime\prime
2}/g_X^2}}J^{Z^{\prime\prime}\mu}Z^{\prime\prime}_{\mu}\notag\\
&-{g_X\over
2\sqrt2}\Big\{\sum_i\sum_j\Big[\Big(\bar\nu_i(DU)^{\dag}_{ij}\Gamma^{\mu}_L
u_j+\bar N_i(DU)^{\dag}_{ij}\Gamma^{\mu}_R u_j\Big)+\Big(\bar
e_i(DU)^{\dag}_{ij}(\Gamma^{\mu}_L+\Gamma^{\mu}_R)d^\prime_j\Big)\Big]
X^{-2/3}_\mu+h.c\Big\}\label{5-1}
\end{align}
where $J^{Z\mu}$ is the neutral current of the standard model
\begin{align}
J^{Z^{\prime}\mu}=-{1\over
4}&\Big\{\tan^2\theta_W\Big[\Big(\bar\nu_i\Gamma^{\mu}_L\nu_i-\bar
e_i\Gamma^{\mu}_L e_i\Big)+\Big(\bar u_i\Gamma^{\mu}_L u_i-\bar
d_i\Gamma^{\mu}_L d_i\Big)-4J^{\mu}_{em}\Big]\notag\\
&+\Big[\Big(\bar N_i\Gamma^{\mu}_R N_i-\bar e_i\Gamma^{\mu}_R
e_i\Big)+\Big(\bar u_i\Gamma^{\mu}_R u_i-\bar d_i\Gamma^{\mu}_R
d_i\Big)\Big]\Big\}\label{5-2}\\
J^{Z^{\prime\prime}\mu}=-{1\over
4}&\Big\{\Big(1-\frac{g^{\prime\prime 2}}{g_X^2}\Big)
\Big[\Big(\bar\nu_i\Gamma^{\mu}_L\nu_i+\bar N_i\Gamma^{\mu}_R
N_i\Big)+2\Big(\bar e_i\gamma^{\mu} e_i-\bar u_i\gamma^{\mu}
u_i-\bar d_i\gamma^{\mu} d_i\Big)\Big]\notag\\
&-{4\over 3}\frac{g^{\prime\prime 2}}{g_X^2}\Big[\bar
u_i\gamma^{\mu} u_i+\bar d_i\gamma^{\mu} d_i\Big]\Big\}\label{5-3}
\end{align}
The gauge group $SU_L(2)\times SU_R(2)\times U_{Y_{1}}(1)\times
SU_X(2)$ is spontaneously broken to the group $SU_L(2)\times
SU_R(2)\times U_{Y}(1)$ by introducing a scalar doublet of $SU_X(2)$
as in the section 2. The vector bosons $m^{\pm 2/3}_{X}$ and
$Z^{\prime\prime}\mu$ acquire the masses
\begin{eqnarray}
m_X^{2}={1\over4}g_X^{2}\text{v}^{\prime2},\quad
m^{2}_{Z^{\prime\prime}}={1\over4}(g_1^{2}+g_X^{2})\text{v}^{\prime2}=
{1\over4}\frac{g_X^{2}\text{v}^{\prime2}}{(1-g^{\prime\prime
2}/g_X^2)}\label{5-4}
\end{eqnarray}
By above symmetry breaking mechanism, the group $SU_L(2)\times
SU_R(2)\times U_{Y}$ is decoupled from the extended gauge group at
mass scale $m_X$.

The group $SU_L(2)\times SU_R(2)\times U_{Y}(1)$ is spontaneously
broken to $U_{em}(1)$ in the standard way \cite{[6]}, by introducing
the scalar doublets
\begin{eqnarray}
\Phi_L :(2,1)_{Y=1},\quad \Phi_R :(1,2)_{Y=1},\quad \Delta_R
:(1,3)_{Y=2}\notag\\
\langle\Phi_L\rangle=\left(\begin{array}{cc}0\\
\text{v}_L\end{array}\right),\quad \langle\Phi_R\rangle=\left(\begin{array}{cc}0\\
\text{v}_R\end{array}\right),\quad \langle\Delta_R\rangle=\left(\begin{array}{cc}0 & \text{V}\\
0&0\end{array}\right)\label{5-5}
\end{eqnarray}
The weak vector bosons, $W^\pm_{L\mu}$, $W^\pm_{R\mu}$,
$Z^\prime_\mu$ acquire masses:
\begin{align}
L_{mass}(W)=&-{1\over 8}\text{v}^2_L\Big[2g^2 W^{-\mu}_L
W^+_{L\mu}+\Big(gW^{\mu}_{3L}-g^\prime
B^\mu\Big)\Big(gW_{3L\mu}-g^\prime B_\mu\Big)\Big]\notag\\
&-{1\over 8}\text{v}^2_R\Big[2g^2 W^{-\mu}_R
W^+_{R\mu}+\Big(gW^{\mu}_{3R}-g^\prime
B^\mu\Big)\Big(gW_{3R\mu}-g^\prime B_\mu\Big)\Big]\notag\\
&-{1\over 8}\text{V}^2\Big[2g^2 W^{-\mu}_R
W^+_{R\mu}\Big]\label{5-6}
\end{align}
where $B_\mu$ is the vector boson associated with $U_Y(1)$, with
coupling constant $g^\prime$. We get
\begin{eqnarray}
m^2_{W_L}&=&{1\over4}g^2\text{v}^2_L,\quad
m^2_{W_R}={1\over4}g^2(\text{v}^2_R+\text{V}^2)\notag\\
m^2_Z&\approx &{1\over4}\frac{g^2}{\cos^2\theta_W}\text{v}^2_L,
\quad m^2_{Z^\prime}\approx
{1\over4}\frac{g^2}{1-\tan^2\theta_W}\text{v}^2_R\label{5-7}
\end{eqnarray}
for $\text{v}^2_L/\text{v}^2_R\ll 1$. We note
\begin{eqnarray}
gW_{3L\mu}-g^\prime B_\mu &=& \frac{g}{\cos\theta_W}Z_\mu
-\frac{g\tan^2\theta_W}{\sqrt{1-\tan^2\theta_W}}Z^\prime_\mu\notag\\
gW_{3R\mu}-g^\prime B_\mu
&=&-\frac{g}{\sqrt{1-\tan^2\theta_W}}Z^\prime_\mu\label{5-8}
\end{eqnarray}
From Eq.(59), it follows that the effective Hamiltonian for the FCNC
involving $e^-$, $\mu^-$, $\tau^-$ is given by
\begin{eqnarray}
\mathcal{H}_{eff}={g^2_X\over m^2_X}\Big[C_{i\alpha}\bar
e_i\Big(\Gamma^\mu_L+\Gamma^\mu_R\Big)d_{\alpha}\Big]\Big[\bar
d_{\beta}\Big(\Gamma_{L\mu}+\Gamma_{R\mu}\Big)\Big]C^*_{j\beta}\label{5-9}
\end{eqnarray}
After Fierz reshuffling:
\begin{eqnarray}
\mathcal{H}_{eff}=\frac{G_X}{\sqrt
2}C_{i\alpha}C^*_{j\beta}\Big[\Big(\bar d_{\beta}\Gamma^\mu_L
d_{\alpha}\Big)\Big(\bar e_i\Gamma_{L\mu} e_j\Big)+2\Big(\bar
d_{\beta}(S-P) d_{\alpha}\Big)\Big(\bar e_i(S-P)
e_j\Big)\Big]\label{5-10}
\end{eqnarray}

In this paper, the effective Hamiltonian given in Eq. (5.10) for
FCNC induced $K,$ $B$ mesons decays is not further discussed.
\section{Summary and conclusion}
The lepton flavor violating decays are strictly forbidden in the
standard model. The gauge group $G=SU_{L}(2)\times U_{Y_1}(1)\times
SU_{X}(2)$ beyond the SM provides a framework to derive the
effective Hamiltonian for FCNC induced decays of $K$ and $B$ mesons
to lepton pairs. The effective coupling constant
$\frac{G_{X}}{\sqrt{2}}=\frac{g^{2}_{X}}{8m^{2}_{X}}$, where
$m_{X}$, the mass of lepto-quark bosons $X^{\pm2/3}_{\mu}$ of
$SU_{X}(2)$ gives the mass scale at which the group $G$ is broken to
$SU_{L}(2)\times U_{Y}(1)$. The effective Hamiltonian gives the
lepton flavor conserving and LF violating decays of the same order.
The upper bound on $(G_{X}/G_{F})$ is obtained from the most
stringent experimental limits on the
$BR(K^{0}_{L}\to\mu^{\mp}e^{\pm})<4.7\times10^{-12}$. Several cases
of pairing three generations of leptons and quarks in the
representation $(2,\bar{2})$ are analyzed. For some cases the upper
bound on $(G_{X}/G_{F})$ is of the order of $(6-9)\times 10^{-6}$
and is compatible with the upper limits on various LF violating
$K$-decays. In particular for these cases, we find the
$B.R(K^{0}_{L}\to\mu^{-}\mu^{+}<(1.95-8.28)\times 10^{-9}$ to be
compared with the experimental value
$B.R(K^{0}_{L}\to\mu^{-}\mu^{+})=(6.84\pm0.11)\times10^{-9}$. It is
tempting to take $(G_{X}/G_{F})$ equal to $(6-9)\times10^{-6}$; this
gives $(g/g_{X})(\frac{m_{X}}{m_{W}})\sim(3-4)\times10^{2}$.

In the SM \cite{[5],[7]},
\begin{eqnarray}
B.R(B^{0}_{s}\to\mu^{-}\mu^{+}=(\frac{\tau_{B^{0}_{s}}}{\hbar})\frac{G^{2}_{F}\alpha^{2}}{16\pi^{3}}
\left|V_{tb}V^{\ast}_{ts}\right|^{2}\left(1-2\frac{m^{2}_{\mu}}{m^2_{B_{s}}}\right)\times
m_{B_{s}}f^{2}_{B_{s}}m^{2}_{\mu}\left|C_{10}\right|^{2}
\end{eqnarray}
Using the experimental values for the masses and
$\tau_{B^{0}_{s}}$,$f_{B_{s}}\approx0.234$,$\left|V_{tb}V^{\ast}_{ts}\right|=3.87\times10^{-2}$
and for the Wilson coefficient $C^{eff}_{10}(m_{b})=-4.13$, one gets
$B.R(B^{0}_{s}\to\mu^{-}\mu^{+})=3.1\times10^{-9}$. More accurate
calculation gives its value $(3.2\pm0.2)\times10^{-9}$(Exp.value
$=(3.2^{+1.5}_{-1.2})\times10^{-9}$).

From Eq. (6.1), we obtain

\begin{eqnarray}
B.R.
(B_s^0\to\tau^+\tau^-)&=&(1-2m_\tau^2/m^2_{B_s})/(1-2m^2\mu/m^2_{B_s})\notag\\
&&\times(m_\tau/m_\mu)^2\text{ B.R. }(B_s^0\to\mu^-\mu^+)\notag\\
&=&(6.9\pm0.4)\times10^{-7}\\
\text{ B.R.
}(B_d^0\to\mu^-\mu^+)&=&\tau_{B_d^0}/\tau_{B_s^0}\frac{f^2_{B_d}m_{B_d}}{f^2_{B_s}m_{B_s}}|\frac{V{td}}{Vts}|^2\approx0.13\times10^{-9}
\end{eqnarray}

For $K_L^0\to\mu^+\mu^-$, Eq. (6.1) is modified to

\begin{eqnarray}
B.R.(K_L^0\to\mu^+\mu^-)&=&2(\tau_{KL}^0/\hbar)\frac{G_F^2}{16\pi^3}|V_{ts}V_{td}^*|^2(1-2\frac{m_\mu^2}{m_K^2})m_K^0f_K^2m_\mu^2\notag\\
&\approx&5.92\times10^{-11}(\frac{C_{10}^{eff}(m_s)}{C_{10}^{eff}(m_b)})^2
\end{eqnarray}

which is much below the experimental value, unless the last factor
on the right hand side of Eq. (6.4) is of order of $10^2$.

For LF violating $B$-decays, the experimental limits on the
branching ratios are not very stringent. However the time integrated
decay rates of $B^{0}_{d,s}\to\ell^{-}\tau^{+}(\tau^{-}\ell^{+})$
due to quantum interference of $B^{0}$ and $\bar{B}^{0}$ is a
promising area to test the model. We find for time integrated decay
rates for $B^{0}_{d,s}\to\ell^{-}\tau^{+}(\tau^{-}\ell^{+})$, the
ratio $\tau^{-}/\tau^{+}=\ell^{+}/\ell{-}\approx 23$ percent, i.e.
depletation of $\tau^{-}$ or $\ell^{+}$ compared to $\tau^{+}$ and
$\ell^{-}$ i.e. violation of charge symmetry.

With more precise experimental data on the LF violating on the
branching ratio for $K^{0}_{L}$ and $B^{0}_{d,s}$ decay to lepton
pairs, it may be possible to test the model.

\end{document}